\RequirePackage{fix-cm}

\documentclass[smallextended]{svjour3}

\smartqed  

\usepackage{graphicx}
\usepackage{xcolor}
\usepackage{url} 
\usepackage{comment}
\usepackage{booktabs}
\usepackage{hyperref}
\usepackage{enumitem}
\usepackage{array}
\usepackage{multirow}
\usepackage{xspace}
\usepackage{listings}
\usepackage{natbib}

\lstdefinelanguage{JavaScript}{
    keywords={break, case, catch, continue, debugger, default, delete, do, else, finally, for, function, if, in, instanceof, new, return, switch, throw, try, typeof, var, let, const, while, with, yield, await, async},
    sensitive=true,
    comment=[l]{//},
    morecomment=[s]{/*}{*/},
    morestring=[b]',
    morestring=[b]"
}
\newcommand{\llama} {Llama 3\xspace}
\newcommand{\gpt} {GPT-4\xspace}
\newcommand{\claude} {Claude 3\xspace}

\journalname{Empirical Software Engineering}

\begin{document}

\title{Do LLMs Consider Security? An Empirical Study on Responses to Programming Questions}

\titlerunning{Do LLMs Consider Security? }   

\author{Amirali Sajadi (corresponding author) \and
        Binh Le                 \and
        Anh Nguyen              \and
        Kostadin Damevski       \and
        Preetha Chatterjee
}

\institute{
    F. Author, S. Author, T. Author, and Fifth Author \at
    College of Computing and Informatics, Drexel University, Philadelphia, PA, USA \\
    \email{amirali.sajadi@drexel.edu, bql23@drexel.edu, adn56@drexel.edu, preetha.chatterjee@drexel.edu}
    \and
    Fourth Author \at
    College of Engineering, Virginia Commonwealth University, Richmond, VA, USA \\
    \email{kdamevski@vcu.edu}
}

\date{Received: date / Accepted: date}

\maketitle

\begin{abstract}

The widespread adoption of conversational LLMs for software development has raised new security concerns regarding the safety of LLM-generated content. Our motivational study outlines ChatGPT's potential in volunteering context-specific information to the developers, promoting safe coding practices. Motivated by this finding, we conduct a study to evaluate the degree of security awareness exhibited by three prominent LLMs: \claude, \gpt, and \llama. We prompt these LLMs with Stack Overflow questions that contain vulnerable code to evaluate whether they merely provide answers to the questions or if they also warn users about the insecure code, thereby demonstrating a degree of security awareness. Further, we assess whether LLM responses provide information about the causes, exploits, and the potential fixes of the vulnerability, to help raise users' awareness.

Our findings show that all three models struggle to accurately detect and warn users about vulnerabilities, achieving a detection rate of only 12.6\% to 40\% across our datasets. We also observe that the LLMs tend to identify certain types of vulnerabilities related to sensitive information exposure and improper input neutralization much more frequently than other types, such as those involving external control of file names or paths. Furthermore, when LLMs do issue security warnings, they often provide more information on the causes, exploits, and fixes of vulnerabilities compared to Stack Overflow responses. Finally, we provide an in-depth discussion on the implications of our findings, and demonstrated a CLI-based prompting tool that can be used to produce more secure LLM responses.

\end{abstract}

\section{Introduction}

Large language Models (LLMs) have become deeply integrated into software engineering workflows, performing tasks such as code generation, summarization, debugging, and addressing queries related to programming~\citep{liu2023better, hou2023large, zheng2023towards, belzner2023large}. In particular, LLM chatbots or \textit{conversational LLMs}, such as OpenAI's GPT~\citep{openai2023chatgpt}, Anthropic's Claude \citep{claude}, and Meta's Llama \citep{llama}, have significantly impacted problem-solving activities by enabling interactive Q\&As \citep{saud2024chatting, das2024investigating, da2024chatgpt}. Developers use them to describe symptoms, provide contextual information, and seek guidance on solutions~\citep{hou2023large}. According to a 2023 survey, 92\% of U.S.-based developers are using various generative models to perform or to automate some of their daily tasks \citep{survey}.

However, the rapid adoption of LLMs by software developers has raised many concerns regarding the security implications of using LLMs.
A recent study found that participants using AI assistants produced code with significantly more vulnerabilities \citep{perry2023users}. Alarmingly, these participants were also more confident in the security of their code, suggesting that AI code assistants can foster a false sense of security, increasing the risk of introducing vulnerabilities into real-world software. Another study found that 32.8\% of Python and 24.5\% of JavaScript code produced by GitHub Copilot are vulnerable \citep{fu2023security}. These vulnerabilities, if exploited, can lead to severe consequences, such as the Log4Shell vulnerability \citep{ibm_log4j_vulnerability}. In 2024 alone, over 34,000 vulnerabilities were reported \citep{CVE_Metrics}, highlighting the increasing frequency and severity of cybersecurity threats that endanger the safety, security, and reliability of software systems.

Beyond generating vulnerable code, using LLMs can impact software security in more intricate and subtle ways. For instance, novice developers may unknowingly input insecure code (copied from Q\&A forums) and ask LLMs to refactor and adapt it to their problem context. Similarly, during debugging, a developer might provide a block of code containing vulnerabilities, such as unsanitized user input in an SQL query, without being aware of its potential security implications. If LLMs fail to identify and address these vulnerabilities, developers may integrate flawed code into their projects, relying on the model without recognizing the potential security risks themselves. To better understand this phenomenon, let us consider the following examples. In a SO question, a developer asks for help to resolve an issue with writing to a file:

\begin{verbatim}
I have a text file with one URL per line, like:

https://www.google.com
https://www.facebook.com
...

The problem is, when I write the resulting URL to a file, I get
anadditional %0A at the end of each line. Can you please explain
to me why is this happening?  I’m using this script to fetch them: 
\end{verbatim}

\begin{lstlisting}[language=Python, caption=]
add = open("manual_list.txt", "r")

for a in add:
    response = requests.get(a, timeout=(2, 5), verify=False)
    fout = open("mylist.txt","a")
    fout.write(response.url+"\n")
    fout.close()
\end{lstlisting}

Here, the developer is focused on removing the extra \%0A characters, which could be resolved using the \texttt{strip()} function i.e., \texttt{requests.get(a.strip(), timeout=(2, 5), verify=False))}. However, they fail to notice a significant security risk posed by the \texttt{verify=False} parameter. Disabling SSL certificate verification is generally considered poor security practice and exposes the application to serious risks, including man-in-the-middle attacks, where an attacker could intercept or manipulate the data sent over HTTPS. If prompted with this question, \gpt, \claude, and \llama all correctly explain the issue with \%0A and suggested using the \texttt{strip()} function to fix it. However, none of the LLMs mentioned the security implications of the \texttt{verify=False} parameter, leaving the developer unaware of the potential vulnerability. By failing to inform the user of this vulnerability, the LLM responses can indirectly reinforce the faulty implementation and, in many cases, further build upon it, perpetuating insecure coding practices.

Several studies have explored the potential risks associated with the use of LLMs and examined concerns regarding the generation of insecure code \citep{pearce2022asleep, siddiq2022securityeval, khoury2023secure, siddiq2024quality}, inaccuracies in vulnerability detection \citep{ullah2024llms, akuthota2023vulnerability, purba2023software, zhou2024large}, and potential misuse for offensive applications, ranging from hardware to user-level attacks \citep{happe2023getting, falade2023decoding}. However, most studies have focused on the security of LLM-generated code, often neglecting the natural language generated by LLMs that plays a critical role in interactive learning and problem-solving~\citep{conversational_prog_blog, xia2023conversational, hao2024empirical}. Additionally, unlike prior research that focuses on detecting vulnerabilities in LLM-generated code or evaluating the capabilities of LLMs when explicitly tasked to detect vulnerabilities, our work investigates the ability of LLMs to \textit{proactively} identify vulnerabilities in user-supplied code. This reflects real-world use cases where developers that rely on LLMs for various tasks inevitably prompt LLMs with vulnerable code. Our study is the first to assess the security awareness of LLMs using the textual information provided by the LLMs alongside or independent of code. We assess not only LLMs' ability to proactively detect vulnerabilities but also their effectiveness in communicating critical information -- such as causes, exploits, and fixes -- to enhance developer understanding and prevent the use of exploitable and insecure code.

We begin with a motivational study using an existing dataset of developer conversations with ChatGPT, investigating the frequency and specificity of vulnerability-related warnings issued by ChatGPT. Mainly, our motivational study aims to understand the way ChatGPT engages with security topics. Initial findings suggest that ChatGPT can offer valuable, context-specific security guidance that encourages safer practices. More importantly, we found instances where the model voluntarily pointed out the security risks. These findings were significant, as they demonstrated ChatGPT’s potential to proactively raise developers' security awareness during development. 
This observation motivated us to dig deeper and perform a systematic examination of the security awareness across three popular LLMs: \claude, \gpt, and \llama. We curated a dataset of 300 Stack Overflow questions that contain vulnerable code. In half of these questions, security vulnerabilities were explicitly noted by the SO participants (Mentions-Dataset); in the other half, vulnerabilities were not identified (Transformed-Dataset). We used these questions as prompts to the LLMs and analyzed their responses, focusing on whether they proactively recognized and addressed security concerns in both code and text-based guidance. Specifically, we investigate the following research questions:

\textit{\textbf{RQ1: }Given insecure code, do LLMs warn developers of the potential security flaws or required security modifications?} 

Through RQ1, we aim to investigate the degree to which LLMs exhibit security awareness. We conduct a qualitative analysis to examine whether LLMs simply provide answers to questions or if they also warn users about the security flaws and suggest potential modifications to improve code security. Our results indicate that LLMs seldom issue security warnings about vulnerabilities unless explicitly prompted to do so. 

\textit{\textbf{RQ2: }In instances where users are reminded of security concerns, are they informed about the causes, potential exploits, and possible fixes of the vulnerabilities?}  \newline
Through RQ2, we aim to determine the types of information included in the LLM security warnings. We qualitatively analyze the information LLMs provided for each vulnerability, specifically checking if they detailed the \textit{causes}, potential \textit{exploits}, and possible \textit{fixes}. We then perform the same analysis on user-provided SO responses containing warnings about insecure code and compare these to the LLM responses across the two datasets. According to our findings, in cases where vulnerabilities are pointed out, LLMs generally offer more information about the causes, exploits, and fixes of the insecurities compared to SO responses.

We discuss the implications of our findings for improving LLM security awareness in Software Engineering (SE) across three key areas: a) Prompt Engineering for SE, b) Integrating LLMs with SE tools, and c) Designing LLMs for SE. By sampling 50 questions from our dataset, we explored various prompting techniques to increase the likelihood of LLMs issuing security warnings. We observed that adding short phrases such as  \textit{``Address security vulnerabilities"} to the prompts showed some effectiveness, although limitations persist. Additionally, we develop a pipeline for integrating outputs from static analysis tools like CodeQL into LLM prompts and find the potential of this method for enhancing security awareness of LLMs. Beyond these immediate solutions, our findings highlight the need for targeted design improvements to address the substantial gaps in LLMs' security awareness, not only in the code they generate but also in the explanations and recommendations they provide. These insights point to key areas for future research, tool development, and LLM evaluations aimed at creating more security-conscious LLMs as programming assistance.

Overall, this paper presents the first study on the security awareness of three popular LLMs in answering programming-related questions. We find that all three LLMs we studied rarely warn developers about security vulnerabilities. We fruther assess how well LLMs can raise developer awareness and encourage the adoption of secure coding practices. We notice that when LLMs do issue security warnings, they often provide more information about the causes, potential exploits, and fixes of the vulnerability, compared to typical SO responses. Observations from this study will inform future designs of tools and evaluation methodologies that aim to make LLM-driven programming more secure. More specifically our paper makes the following contributions:

\begin{itemize}[label=\textbullet]
    \item A motivational study that examines the naturally occurring security-related conversations between developers and ChatGPT and outlines the ability of LLMs for proactively warning users about security.

    \item A benchmark for evaluating: (a) LLMs' capabilities in \textit{proactively detecting vulnerabilities} in real-world user-supplied code, and (b) issuing warnings with critical information—such as \textit{causes}, \textit{exploits}, and \textit{fixes}—to enhance developer understanding and mitigate integration of insecure code in existing code-bases.

    \item Preliminary results to demonstrate the opportunities for adapting simple and practical prompt engineering techniques to enhance the security of the LLM responses to programming questions.

    \item A CLI-based tool that analyzes developer queries and integrates CodeQL outputs to generate prompts that result in significantly safer LLM responses.
\end{itemize}

\section{Motivational Study}
\label{motivational_section}

\begin{figure}[t]
  \centering
  \includegraphics[width=\linewidth]{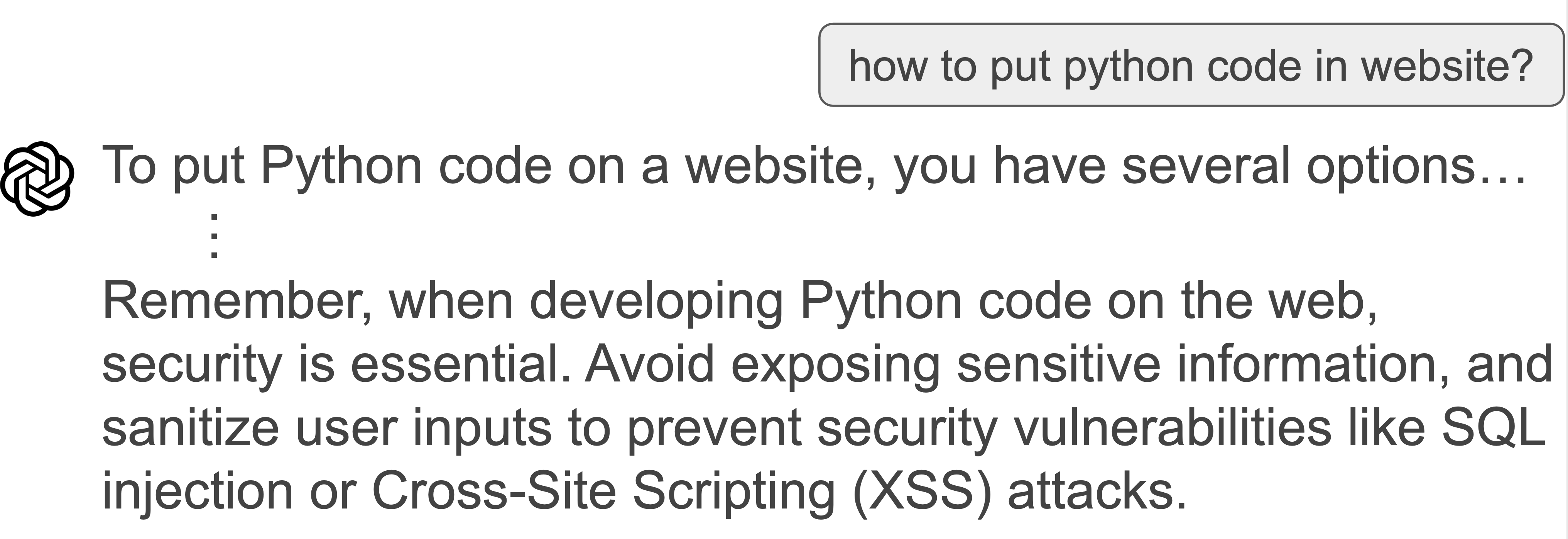}
  \caption{DevGPT: ChatGPT Reminding User about the Security of the Code.}
  \label{fig:motivational_example}
\end{figure}

The 2023 JetBrains survey, based on responses from 26k developers across 196 countries, reveals that 77\% (i.e., approximately three in four developers) use ChatGPT~\citep{jetbrains_survey}. Given its widespread adoption, we conducted a exploratory study to explore whether security considerations naturally emerge in developer conversations with ChatGPT. This motivational study serves as an initial gauge to determine if LLMs engage with security topics at all, helping us assess the feasibility of a larger, more comprehensive study. By examining existing ChatGPT conversations with explicit security mentions, we aimed to understand the frequency and depth of security-related advice offered by the model. This approach allowed us to examine patterns in ChatGPT’s security guidance, including the initiating party of these discussions and the specific topics addressed. Fig \ref{fig:motivational_example} illustrates part of a conversation in which ChatGPT issues a security warning, highlighting the security implications of the suggested code.

\noindent
\textit{\textbf{Dataset: }} In May 2023, OpenAI introduced a feature that allows users to share their conversations with ChatGPT through dedicated links~\citep{openai2023chatgpt}. Using this feature, \cite{xiao2023devgpt} collected all the 3,794 developer-ChatGPT conversations publicly shared on GitHub and Hacker News until October 2023, forming the DevGPT dataset. To identify conversations that mention security, we performed a keyword search on the text of these ChatGPT conversations. As keywords, we used the SO tags related to security proposed by \cite{yang2016security}, such as ``security", ``web-security", ``sql-injection", and ``xss". To ensure comprehensiveness of our data selection, we used both hyphened and non-hyphened variations of the keywords, such as ``web-security" and ``web security", where applicable. This filtering step resulted in a subset of the DevGPT dataset containing 233 conversations. We excluded 13 out of these 233 conversations that included security-related keywords but are predominantly in languages other than English, resulting in 220 conversations.

Although DevGPT primarily contains developer-ChatGPT interactions, not all conversations are guaranteed to relate to software engineering (e.g., programming concepts, debugging, or code generation). Therefore, we conducted a manual check to filter out such non-SE conversations. In addition, as part of this manual step, two authors validated the performance of our keyword search by distinguishing conversations that discuss security from those that merely included security-related keywords without actually addressing security topics, i.e., false positives. The Cohen's kappa agreement~\cite{mchugh2012interrater} for classifying conversations as security-related or non-security-related achieved the strong score of 0.799, after which, the authors resolved all conflicts and finalized the data through discussion. Finally, we identified 102 technical conversations with ChatGPT mentioning software security, shared across various data sources, including code files (45), issue threads (33), Hacker News (11), pull requests (9), and GitHub commits (4). 

\noindent
\textit{\textbf{Procedure: }} We conducted a qualitative analysis to investigate the role of security in the 102 developer-LLM conversations with mentions of security. We identified which party (Developer or ChatGPT) brings up the security considerations, examined the types of available information, and identified the type of vulnerability mentioned in each conversation. Two authors of this paper manually annotated the dataset (annotation instructions with examples are included in our replication package). The analysis was performed in an iterative approach consisting of multiple sessions. To calculate inter-rater agreement, we used Cohen's Kappa coefficient. This process resulted in substantial agreement values (i.e., $>$ 0.6): 0.88 for \textit{types of information}, 0.66 for \textit{types of vulnerability}, and 0.62 for \textit{initiating party}.

\underline{Types of Information:} To examine the types of information in the ChatGPT conversations, we used the \cite{pan2021automating} taxonomy for for categorizing information in developer chats. Developer chats (e.g., Slack, Discord) closely align with developer-LLM conversations since both support dynamic, rapid, and iterative information exchange~\citep{Chatterjee2019}. Consequently, we believe this taxonomy is well-suited for our study. The information categories in the taxonomy are listed in Table \ref{tab:motivational_overall_topic}. 

\begin{table}
    \centering
    \caption{Distribution of Types of Information in DevGPT Conversations with Security Mentions.}
    \label{tab:motivational_overall_topic}
    \begin{tabular}{p{0.68\linewidth}  p{0.2\linewidth}}
        \toprule
        \textbf{Information type}  & \textbf{Instances} \\
        \midrule
        Technical Discussion  & 49\\
        Programming Information & 31\\
        Programming Problems  & 10\\
        General Information & 5\\
        Documentation Information  & 4\\
        Library Problems &  3\\
        \bottomrule
    \end{tabular}
\end{table}

\underline{Types of Vulnerability:}  The understand the types of vulnerabilities that were discussed, we used a taxonomy proposed by \cite{russo2019summarizing} and manually identified the relevant category for each conversation. We chose Russo et al.'s taxonomy because it provided a predetermined set of vulnerability types, allowing us to present an overview without needing to determine the level of granularity in CWEs for each vulnerability. Russo et al. categorize vulnerabilities into ten distinct types, providing a broad yet concise overview suitable for our analysis.

Additionally, we took note of whether ChatGPT makes broad mentions of security, or if it offers detailed information about security. We define a specific mention as a mention that contains any form of implementation details. An example of specific mention could be a conversation in which ChatGPT warns the user about the possibility of an SQL injection based on the prompt's code and offers ways to preventing it. On the other hand, we define a broad mention as a mention that lacks any implementation details, such as the sentence ``\textit{It's important to consider additional security measures when dealing with file uploads...}". 

\underline{Initiating Party:} We also determined whether it is ChatGPT or the user who first introduces the topic of security into the conversation. Specifically, we identified instances where ChatGPT provided security-related information or where the user directly asked about security aspects of development. In cases where the question inherently involves security but lacks explicit security mentions in the user's prompt, we do not attribute the initiation of the security discussion to either party.

\noindent
\textit{\textbf{Findings: }}
 Table \ref{tab:motivational_overall_topic} illustrates the distribution of the \textbf{types of information} available in all 102 conversations in our dataset. The most common type of information is \textit{Technical Discussion}, which relates to conceptual conversations about software engineering. The frequent occurrences of security mentions in \textit{Technical Discussions} points to the fact that security is more often included when conversations revolve around higher level issues related to software development. In many of these instances, ChatGPT informs the users about the good practices with regard to security, making statements such as: ``\textit{Always ensure that you're following the correct steps as mentioned in the WordPress.com OAuth2 documentation, and handle the tokens securely, keeping them out of URLs whenever possible to maintain security}". \textit{Programming Information}, i.e., conversations in which the user is mostly trying to get the LLM to generate code or to explain programming concepts, and \textit{Programming Problems}, i.e., conversations in which the user is mostly trying to resolve programming problems, are the second and third most common types of conversations that lead to notions of security. In these cases, we often see ChatGPT trying to inform the user about the security concerns in their code or even the code generated by ChatGPT itself as well as the potential fixes for these concerns. Security, however, is rarely brought up in conversations about the user's problems with specific libraries or when the users are attempting to retrieve any general or documentation related information. \textit{General Information}, i.e., discussions that are not closely related to the project itself, such as the best choice of IDE or job-hunting experiences, \textit{Documentation Information}, and \textit{Library Problems} are the least frequently observed types of information in conversations.

Table \ref{tab:motivational_taxonomy} provides an overview of the \textbf{types of vulnerabilities}. Notably, ``\textit{Authentication bypass or Improper Authorization}" and ``\textit{Cross-Site Scripting or HTML Injection}" are the most frequently discussed types of vulnerability, while others have much fewer mentions. Further, some types of vulnerabilities such as ``\textit{Buffer/Stack/Heap/Integer Overflow, Format String and Off-by-One}" were not present in any conversation. We also observed that 54 conversations include broad mentions of security and the other 48 conversations include specific mentions. These results indicate that in many instances ChatGPT makes statements about security without pointing out a specific vulnerability.

\begin{table*}
    \centering
    \caption{Summary of Vulnerability Categories Mentioned by ChatGPT Throughout Conversations.}
    \label{tab:motivational_taxonomy}
    \begin{tabular}{
            >{\raggedright\arraybackslash}p{0.2\linewidth}
            >{\raggedright\arraybackslash}p{0.24\linewidth}
            >{\raggedright\arraybackslash}p{0.1\linewidth}
            >{\raggedright\arraybackslash}p{0.32\linewidth}}        
        \toprule
        \textbf{Vulnerability Category} & \textbf{Description} & \textbf{Instances} & \textbf{Example} \\
        \midrule
        \textbf{Authentication bypass or Improper Authorization} & Allowing attacker to bypass required authentication or not performing required authentication checks & 18  & \textit{authorization code should be unique and temporary for each user session...each code can only be exchanged once for security reasons.}\\
        \textbf{Cross-Site Scripting or HTML Injection} & Allowing attacker to execute arbitrary code in the web browser and stealing cookie-based credentials. & 16  & \textit{This means that the cookie can only be accessed via HTTP requests and not through client-side scripts, which is a good security measure to prevent XSS attacks.}\\
        \textbf{SQL Injection} & Not properly sanitizing user input before using them in SQL queries. & 5  & \textit{Prepared statements are efficient and safe against SQL injection attacks.}\\
        \textbf{Information Disclosure and/or Arbitrary File Read} & Allowing attacker to get access to information and files. & 3  & \textit{Avoid hardcoding them in your Python script. Instead, use environment variables or AWS profiles.}\\
        \textbf{Directory Traversal} & Allowing attacker to gain read access to arbitrary file content. & 2  & \textit{Exposing the entire node\_modules directory publicly is generally a bad idea, due to the potential security risks and unnecessary exposure of dependencies.}\\
        \textbf{Remote Code Execution} & Allowing attacker to execute arbitrary code within the affected application, potentially leading to unauthorized access or a privilege escalation. & 1  & \textit{The vm2 library is a sandbox that can run untrusted code securely. It's built on top of the Node.js vm module and adds additional security.}\\
        \bottomrule
    \end{tabular}
\end{table*}

Our findings about the \textbf{initiating party} show that in 69 out of the 102 cases, ChatGPT is the one who first mentioned security. In 14 out of those 69 conversations, the user followed up on this mention of security and further discussed the topic, while in 49 conversations the users did not directly follow up on the mention of security and in 6 instances the conversation did not continue by the user. Further, out of the 69 instances where ChatGPT first brought up the topic of security, 42 mentions were broad, while 24 were specific. In the remaining 33 of the 102 conversations, it was the user who first mentioned security. For example, one user said ``I have this class for generating user tokens... \textless{}code snippet\textgreater{}... What has better security, my class or using SHA-256?"

Overall, through this analysis several key observations have emerged. The existence of 102 instances with mentions of security within DevGPT dataset indicates a degree of emphasis on security. Further, in 69 of these 102 conversations, ChatGPT volunteered the security information, without a direct request. 49 of these 69 instances were broad mentions e.g., ``\textit{Lastly, always keep the user's privacy and security in mind.}", while 24 contained implementation details tailored to the user's specific use-case. In addition, 14 of these 69 mentions led to further inquiries about security by the users. For instance, in one conversation, ChatGPT reminded the user about the importance of using prepared statements in order to avoid SQL injection, at which point, the user asked ChatGPT to rewrite this code with prepared statement.

Given our observations, this \textit{empirical exploration} of real-world developer conversations with ChatGPT highlights its potential to proactively inform developers about security vulnerabilities and provide context-specific solutions for writing secure programs. Although infrequent, such proactive contributions promote security awareness, helping developers avoid insecure practices and prevent them from building upon vulnerable code. These findings highlighted the importance of systematically studying this behavior. Consequently, we designed our main study to evaluate the extent and consistency of this behavior across three popular LLMs, under controlled scenarios.

\section{Methodology}

Building upon our findings from the motivational study, we designed an \textit{experimental study} to systematically evaluate the security awareness of three prominent LLMs, \claude, \gpt, and \llama. Our goal is to determine whether LLMs can proactively detect vulnerable code in prompt inputs, and how consistently they issue security warnings  and relevant vulnerability-related information to the users. To this end, we first collected Stack Overflow (SO) questions containing vulnerable code snippets, which we then used as prompts for the LLMs. We qualitatively analyzed the LLM responses to address our research questions.

\begin{figure*}[t]
  \centering
  \includegraphics[width=\linewidth]{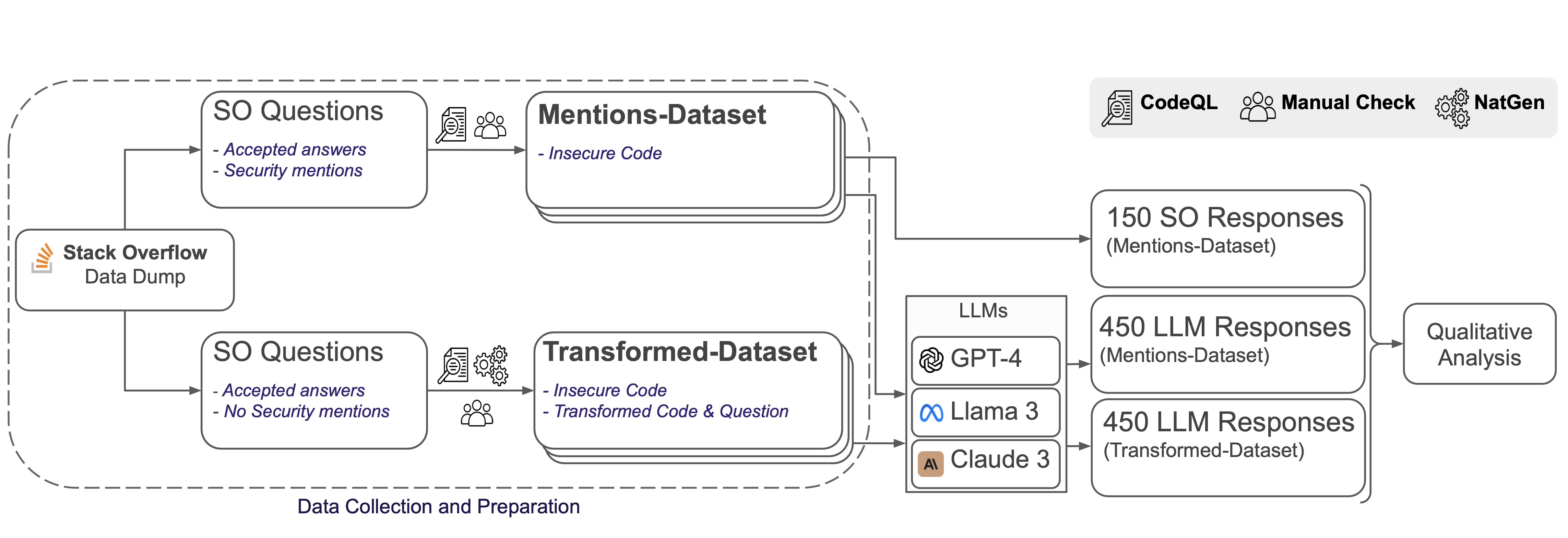}
  \caption{Overview of our study methodology.}
  \label{fig:methodology}
\end{figure*}

By examining whether an LLM issues a warning about the security of the code in the SO question, we evaluate the LLM's security awareness. When an LLM not only answers the question but also highlights the security issue in the code, it demonstrates a high level of security awareness. Conversely, if the LLM addresses the question without mentioning security, it demonstrates a low level of security awareness. This premise forms the basis for our assessment of LLMs' security awareness in answering developers' questions. Figure ~\ref{fig:methodology} provides an overview of our approach, which we discuss in detail next. 

\subsection{Dataset Collection and Refactoring}

\noindent
\textbf{Data Collection.}

\begin{figure}[t]
  \centering
  \includegraphics[width=\linewidth]{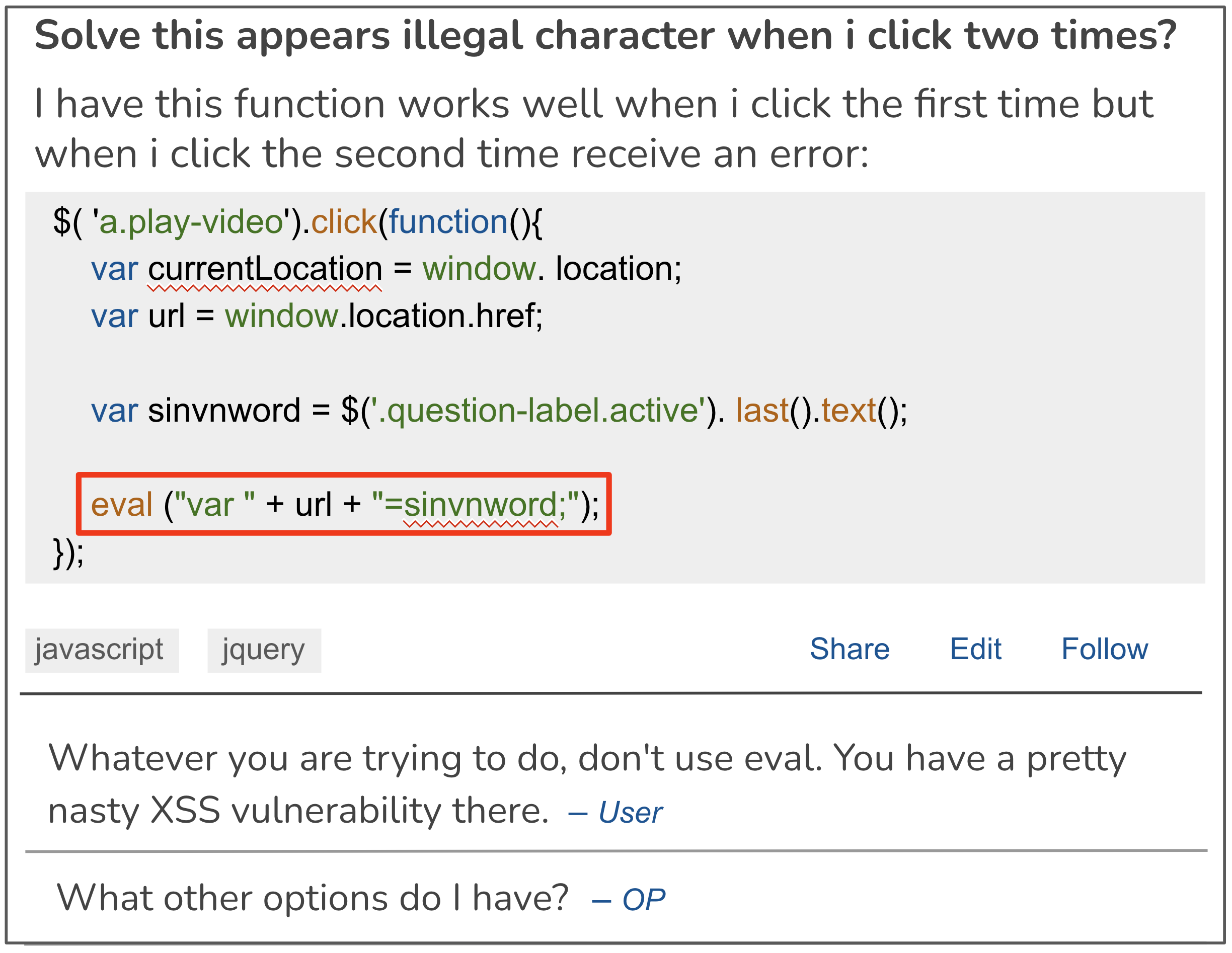}
  \caption{Example of a SO Question Containing Vulnerable Code (Highlighted Red), and a User Response (Comment), Pointing Out the Vulnerability.}
  \label{fig:example_insecure_so}
\end{figure}

We collect a dataset of SO questions with insecure code but without any explicit mention of security within the questions themselves. Figure \ref{fig:example_insecure_so} shows an example of such a question, where the JavaScript code uses the \textit{eval} function to dynamically create a variable based on user input, introducing a significant risk of code injection or a cross-site scripting (XSS) attack. The developer posting the question seems unaware of the related security issue; however, one of the respondents points this out in a comment \textit{``You have a pretty nasty XSS vulnerability here"}.

However, not all SO questions containing vulnerable code receive responses that point out the security implications of the code. In fact, earlier studies highlight the concerns with developers copy-pasting insecure code from SO posts \citep{fischer2017stack}. Therefore, we aim to create two datasets for our study: (a) \textit{\textbf{Mentions-Dataset:}} a set of questions that received responses (either answers or comments) mentioning and/or addressing the security concerns in the question's code (as shown in Figure \ref{fig:example_insecure_so}), (b) \textit{\textbf{Transformed-Dataset:}} a set of questions that despite containing vulnerable code, do not receive such responses. The code in this dataset has been transformed (Later detailed in this section) to ensure that it is not easily recognizable for the LLMs.

The Mentions-Dataset represents a best-case scenario where the vulnerability has already been highlighted by the community, and the code appears exactly as it was originally posted on SO. This allows us to observe how LLMs respond to questions that both exist in their training data and include explicit security warnings. In contrast, the Transformed-Dataset intentionally excludes any mention of security in the responses and applies code refactoring to reduce the chance of data leakage from the LLM’s training data. By comparing these two datasets, we can examine LLM behavior across a spectrum of possible real-world situations: from code that has been explicitly marked as vulnerable to code that has not been associated with any vulnerabilities by the SO responses. This contrast helps us assess not only how well LLMs leverage familiar content (Mentions-Dataset) but also their ability to generalize to insecure code they may not have associated with vulnerabilities before (Transformed-Dataset).

To collect these datasets, we used the SO data dump \citep{stackoverflow2023} from March 2015 to March 2024, which contains over 10 million SO questions. First, we selected questions that contain tags ``python" or ``javascript" to focus on these widely-used programming languages. According to the \cite{stackoverflow2024} Developer Survey, both JavaScript and Python are listed in the top 3 most commonly used languages, making our results more representative of common real-world scenarios. We excluded questions that did not have accepted answers, resulting in a total of 4.89 million questions. To ensure that there were no mentions of security in the natural language section of the questions, we performed keyword searches using the same keywords from Section \ref{motivational_section}. We discarded questions containing any of those keywords, leaving us with 4.77 million questions. Next, we discarded the questions that did not contain code snippets and analyzed the code in remaining questions using CodeQL \citep{GitHubCodeQL}, a static analysis tool often used to identify security vulnerabilities \citep{hamer2024just, siddiq2024quality, pearce2022asleep}. As CodeQL does not require access to the entire codebase, it effectively detected vulnerabilities directly within the code snippets. This step resulted in the detection of 4935 questions with insecure code snippets.

We also made sure that each question in Mentions-Dataset has received at least one answer or comment containing a security-related keyword. Similarly, we ensured that Transformed-Dataset only consists of questions that have no security-related keywords in their answers or comments. 
At each step of the filtering process, three authors manually checked the datasets to ensure their integrity. By  inspecting all the instances in the subset of Mentions-Dataset, we identified 150 unique SO questions i.e., 75 Python and 75 JavaScript. Next, we selected the same number of questions form the Transformed-Dataset, resulting in a total of 300 questions.

\noindent
\textbf{Data Refactoring.}
One concern when testing the performance of LLMs on these questions is the possibility that the LLM may have been trained on them \citep{AIStackExchange2023}. This is a deliberate attribute of the Mentions-Dataset, where prior exposure to the questions, their code, and associated responses is expected and intended. The purpose of this dataset is to examine LLM behavior in scenarios where data leakage is likely, allowing us to evaluate how well LLMs leverage familiar content and existing security discussions.

In contrast, to examine the LLMs’ ability to generalize to new or unseen inputs, we applied code refactoring techniques to the Transformed-Dataset in line with prior research \citep{steenhoek2024comprehensive}. Even when explicit mentions of vulnerabilities are absent in SO responses, similar or identical vulnerable code could exist on other platforms, such as GitHub repositories or coding tutorials, and these resources can potentially be included in the LLM’s training data. Therefore, to minimize the possibility of data leakage and maximize the generalizability of our evaluation, we applied meaning-preserving transformations to the vulnerable code snippets. These transformations help reduce input similarity and overlap with the LLM’s training data. These transformations, consistent with recent research on data leakage \citep{ramos2024large}, reduce \textit{n-gram overlap} and increase \textit{Negative Log Likelihood (NLL)} through techniques like \textit{variable renaming}, making the input code less ``natural" to the model. This mitigation of memorization enables a more robust evaluation of the LLMs’ ability to generalize beyond their training data.

We used the NatGen tool \citep{chakraborty2022natgen} to refactor the code while preserving their vulnerabilities. We applied multiple refactoring techniques, such as \textit{dead-code insertion} and \textit{loop transformations} to each code snippet, wherever applicable. For questions containing multiple code snippets, we ensured that all snippets were subjected to transformations. \textit{Variable renaming}, a transformation where the actual variable names are changed to dummy names, was applied only to the code snippet containing the security vulnerability. This selective application was necessary to maintain consistency among variable names across different code snippets. For example, if a variable name is changed to \texttt{VAR\_0}, this change must be reflected, uniformly, in all code snippets within the same question. To ensure this consistency, after performing the variable renaming transformation on the vulnerable code snippet, we automatically updated the variable names throughout all the code snippets to match the new names. 

Additionally, we ensured that the renamed variables were updated in the natural language parts of the questions. Given that various parts of the code snippets, including variable names, may be directly referenced in the text, a manual verification was necessary to ensure coherence in the data. One of the authors manually inspected and adjusted all 150  questions in Transformed-Dataset, ensuring  consistency between the code snippets and the natural language text. 

\subsection{Experimental Setup}

After finalizing the datasets, we used the SO questions as prompts for the three LLMs. Each SO post served as a single prompt, and the responses generated by each LLM were subsequently collected for analysis. For this study, we utilized three prominent LLMs: \llama with 70 billion parameters, \claude Opus, and \gpt Turbo.

\llama was run locally on a server, while \gpt Turbo and \claude  Opus were accessed via their official APIs. To ensure more deterministic answers, all models were configured to generate content with a temperature of 0.1, in line with previous studies \citep{savelka2023can, siddiq2024using}. Additionally, the top-p parameter was left at its default value, as recommended by the \cite{openaidoc} documentation.

\noindent
\textit{\textbf{RQ1: Given insecure code, do LLMs warn developers of the potential security flaws or required security modifications?}} To address RQ1, we conducted a qualitative analysis of the responses generated by each LLM for all 300 questions across our two datasets. Thus, this process involved analyzing a total of (300*3) = 900 LLM-generated responses. 

In order to gain deeper insights into the LLM responses and behaviors, we analyzed and compared the performance of the three LLMs in terms of their ability to identify the vulnerabilities and subsequently inform the users about them. Three authors manually inspected all 900 LLM responses, and evaluated whether each response contains a warning about the security vulnerability or if it only answers the main question.

Additionally, we performed a comparative analysis of the LLMs' performances on questions from Mentions-Dataset against those from Transformed-Dataset. This comparison aimed to highlight any potential influence of the LLMs' training data on their performance. By examining these differences, we sought to understand the extent to which prior exposure to similar questions may affect the LLMs' ability to recognize vulnerabilities in new inputs and warn the users about them.

\noindent
\textit{\textbf{RQ2: In instances where users are reminded of security concerns, are they informed about the causes, potential exploits, and possible fixes of the vulnerabilities?}} 

\noindent
To answer RQ2, we qualitatively analyzed all 900 LLM-generated, and 150 user-provided responses that contained security warnings. Our goal was to evaluate how effectively each response can increase user's awareness of the security implications in their code. For instance, a security mention could range from a brief statement, such as \textit{``this is a security loophole,"} to a detailed explanation, such as, \textit{``Adding verify=False like you did 'fixes' the problem by adding even more insecurity, i.e., it skips the validation of the certificate. See \underline{here} for how such situations can be properly fixed. I recommend..."} The second example can be much more effective in communicating the security concerns to the user and leading them in the right direction.

Numerous studies have explored the types of information developers find valuable in Q\&A-style conversations~\citep{ Treude:2011:PAA:1985793.1985907, lill2024helpfulness, Rosen2015WhatAM, 7335404, 8493285, Imran_clarification}. We adapted the information categories identified by \cite{ChatterjeeJSS} for finding help with programming errors on SO, such as \textit{symptom cause}, \textit{posted solution}, and \textit{solution justification}. We modified these categories to focus on information that  help users better understand the security of their code, as follows:

\begin{itemize}[leftmargin=*]
    \item \textbf{Cause:} The underlying cause of the vulnerability e.g., using outdated hashing algorithms or hard-coding passwords.
    \item \textbf{Exploits:} The ways vulnerable code could be exploited e.g., SQL injection, XSS attacks, or remote code execution.
    \item \textbf{Fixes:} The possible measures to mitigate the security issues e.g., sanitizing user input in order to prevent SQL injection.
\end{itemize}

By identifying the types of information at a more granular level, we aim to determine whether the security warnings provided by the LLMs are broad and ambiguous or if they offer substantial information that enhances user's understanding of the security issues, thereby promoting safer coding practices.

In our evaluation, we compared the performance of all three LLMs to determine how effectively each LLM informs the users about vulnerabilities. We also compared the responses of the LLMs with responses of the SO users in Mentions-Dataset i.e., the cases where both the SO responses and the LLM answers provide security warning. This comparison highlights the differences in the level of detail and helpfulness of the security information provided by humans and LLMs.

Finally, we examined the performance of the LLMs in providing security information in Mentions-Dataset compared to Transformed-Dataset. This comparison aimed to identify whether the presence of security mentions in SO responses of Mentions-Dataset influenced the LLMs to provide more comprehensive answers compared to Transformed-Dataset.

\section{Results}

\subsection{\textbf{RQ1: Given insecure code, do LLMs warn developers of the potential security flaws or required security modifications?}}
\label{rq1}

In Table~\ref{tab:mentions_label} presents the number and percentage of questions in which LLMs warned users about the security vulnerabilities of their code, across the two datasets. Overall, the LLMs provide security warnings for only a small fraction of questions; none of the LLMs warned developers about vulnerabilities in more than 40\% of questions. Further, the performance of all LLMs showed noticeable differences between Mentions-Dataset and Transformed-Dataset. In Mentions-Dataset, where security concerns were present in SO answers, the LLMs were also more likely to identify the security flaws and issue warnings about them. On the other hand, all LLMs faced more difficulties detecting the vulnerabilities contained in the questions from the Transformed-Dataset. On average, the three LLMs, detected 30\% of the vulnerabilities for the questions from the Mentions-Dataset, while detecting only 13.7\% of the vulnerabilities in the Transformed-Dataset. As mentioned before, the varying performance of LLMs on two datasets can be due to their reliance on their training data, pointing out their limitations in generalizing beyond their training data. Additionally, despite the relatively poor performance of all three models, the gap between \gpt and the other two models remains significant. \gpt consistently outperformed the other models across both datasets, with a particularly notable difference in the Mentions-Dataset: 18\% higher detection rate compared to \claude and 12\% higher compared to \llama.

\begin{table}
    \centering
    \caption{Number of Questions Where Vulnerable Code Resulted in a Security Warning by LLMs in Each Dataset.}
    \label{tab:mentions_label}
    \begin{tabular}{p{0.2\linewidth}  p{0.2\linewidth} p{.2\linewidth} p{0.2\linewidth}}
        \toprule
        & \textbf{\llama} & \textbf{\claude} & \textbf{\gpt} \\ 
        \midrule
            \textbf{Mentions-Dataset} & 42/150 (28\%) & 33/150 (22\%) & 60/150 (40\%) \\ 
            \textbf{Transformed-Dataset} & 19/150 (12.6\%) & 18/150 (12\%) & 25/150 (16.6\%) \\
        \bottomrule
    \end{tabular}
\end{table}

\begin{figure}
    \centering
    \includegraphics[width=0.8\linewidth]{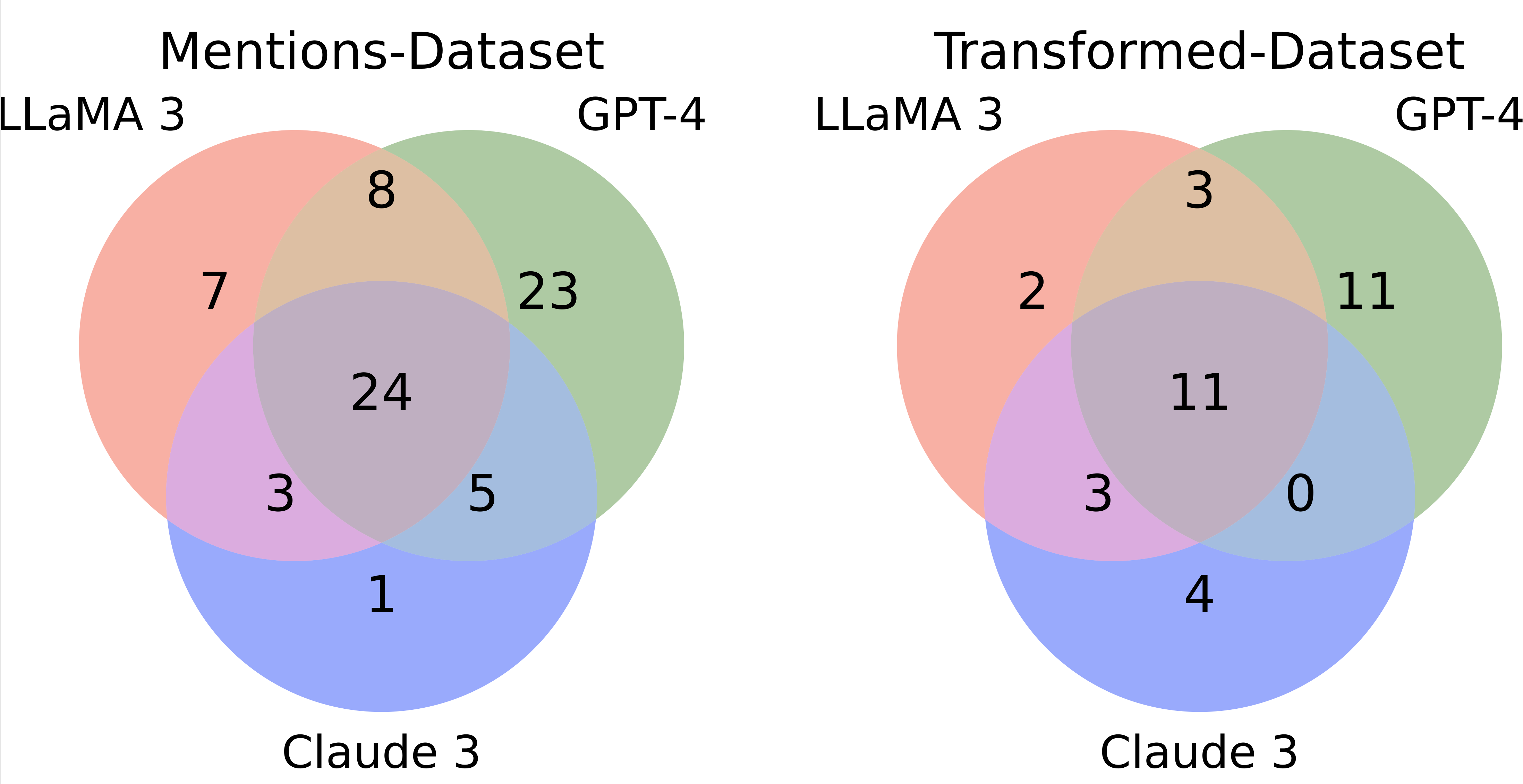}
    \caption{Overlap of the Identified Vulnerable Questions by the Three LLMs}
    \label{fig:overlap}
\end{figure}

Figure~\ref{fig:overlap} illustrates the overlap of questions where the LLMs successfully identified vulnerabilities and informed users about them. The security concerns in 24 questions from the Mentions-Dataset and 11 questions from the Transformed-Dataset were pointed out by all three LLMs. \gpt issued security warnings for 23 vulnerabilities in the Mentions-Dataset and 11 in the Transformed-Dataset for questions where \llama and \claude only provided direct answers. 
Specifically, \llama issued unique security warnings for 7 questions in the Mentions-Dataset and 2 in the Transformed-Dataset. In contrast, \claude  issued only 1 warning in the Mentions-Dataset and 4 in the Transformed-Dataset. The performance overlap between \claude  and \gpt in the Mentions-Dataset (5 questions), is more significant than the overlap between \claude  and \llama (3 questions). However, in the Transformed-Dataset, \claude 's performance is more similar to \llama; both models identified vulnerabilities in 3 questions that \gpt did not detect. There were no questions where \claude and \gpt both identified vulnerabilities that \llama  missed.

Table~\ref{tab:detected_vulnerabilities} categorizes the most frequently detected vulnerabilities by CWE. These CWEs only include those that appeared in five or more questions from the Mentions-Dataset and two or more questions from the Transformed-Dataset. Note that the Transformed-Dataset contains a wider variety of vulnerabilities, resulting in fewer instances of each vulnerability CWE. Furthermore, the CWE mappings were derived from the output of CodeQL analysis. CodeQL queries are tied to one or more CWEs, explaining why some questions are associated with multiple CWEs. For example, the \textit{command-line-injection} query in CodeQL detects vulnerabilities that affect command line interfaces or OS-level operations. Such vulnerabilities typically arise from inadequate input validation and sanitization. Consequently, both CWE-78 (OS Command Injection) and CWE-88 (Argument Injection) are linked to this query. Overall all LLMs exhibit a consistent detection pattern for certain vulnerabilities. As shown in  Table~\ref{tab:detected_vulnerabilities}, CWEs such as CWE-532 (Insertion of Sensitive Information in Log File), CWE-321 (Use of Hard-coded Cryptographic Key), and CWE-798 (Use of Hard-coded Credentials) are among the most frequently detected vulnerabilities, being mentioned in 31.4\%, 29.6\%, and 29.6\% of the LLM answers. On the other hand, among the vulnerabilities that appeared in more than 5 and 2 questions from the Mentions and the Transformed-Dataset, CWE-400 (Uncontrolled Resource Consumption) was only detected in 11.11\% of the cases, while CWE-489 (Active Debug Code) was detected in 14.1\% of the relevant instances. Additionally, many other CWEs such as CWE-1333 (Inefficient Regular Expression Complexity), CWE-22 (Path Traversal), and CWE-99 (Resource Injection), which made more rare appearances in our datasets, went either completely undetected or were identified at most once. Furthermore, the types of vulnerabilities detected by the LLMs, did not significantly vary across the two datasets. As expected, prior exposure to similar SO posts in the training data (i.e., Mentions-Dataset) makes the LLMs more likely to communicate security concerns to the users. However, the LLMs' tendencies to point out certain vulnerabilities over others remained consistent across our datasets.

\begin{table*}
    \centering
    \caption{Frequently identified CWEs in LLM responses to questions form the Mentions-Dataset (M), the Transformed-Dataset (T), and their Combination (MT). CWE descriptions: \textbf{532} - Insertion of Sensitive Information in Log File; \textbf{321} - Use of Hard-coded Cryptographic Key; \textbf{798} - Use of Hard-coded Credentials; \textbf{259} - Use of Hard-coded Password; \textbf{312} - Cleartext Storage of Sensitive Information; \textbf{359} - Exposure of Private Personal Information; \textbf{79} - Cross-site Scripting; \textbf{116} - Improper Encoding or Escaping of Output.}
    
    \label{tab:detected_vulnerabilities}
    \begin{tabular}{p{0.06\linewidth} p{0.038\linewidth} p{0.038\linewidth} p{0.038\linewidth} p{0.038\linewidth} p{0.038\linewidth} p{0.038\linewidth} p{0.038\linewidth} p{0.038\linewidth} p{0.045\linewidth} p{0.042\linewidth} p{0.042\linewidth} p{0.04\linewidth}}
        \toprule
        \textbf{CWE} & \multicolumn{3}{c}{\textbf{\llama}} & \multicolumn{3}{c}{\textbf{\claude}} & \multicolumn{3}{c}{\textbf{\gpt}} & \multicolumn{3}{c}{\textbf{All Models (Avg.)}} \\ 
        \cmidrule(lr){2-4} \cmidrule(lr){5-7} \cmidrule(lr){8-10} \cmidrule(lr){11-13}
        & \textbf{M} & \textbf{T} & \textbf{MT} & \textbf{M} & \textbf{T} & \textbf{MT} & \textbf{M} & \textbf{T} & \textbf{MT} & \textbf{M} & \textbf{T} & \textbf{MT} \\
        \midrule
        \textbf{532}& 10/29 & 0/6 & 10/35 & 7/29 & 0/6 & 7/35 & 15/29 & 1/6 & 16/35 & 36.5\%& 5.0\%& 31.4\%\\
        \textbf{321}& 6/20 & 2/7 & 8/27 & 5/20 & 2/7 & 7/27 & 7/20 & 2/7 & 9/27 & 30.0\%& 28.5\%& 29.6\%\\
        \textbf{798}& 6/20 & 2/7 & 8/27 & 5/20 & 2/7 & 7/27 & 7/20 & 2/7 & 9/27 & 30.0\%& 28.5\%& 29.6\%\\
        \textbf{259}& 6/20 & 2/7 & 8/27 & 5/20 & 2/7 & 7/27 & 7/20& 2/7 & 9/27 & 30.0\%& 28.5\%& 29.6\%\\
        \textbf{312}& 14/45 & 1/12 & 15/57 & 8/45 & 1/12 & 9/57 & 19/45 & 2/12 & 21/57 & 28.3\%& 10.8\%& 26.3\%\\
        \textbf{359}& 14/44 & 0/11 & 14/55 & 8/44 & 0/11 & 8/55 & 19/44 & 2/11 & 21/55 & 35.4\%& 5.4\%& 26.0\%\\
        \textbf{79}& 7/38 & 4/18 & 11/56 & 7/38 & 2/18 & 9/56 & 18/38 & 3/18 & 21/56 & 27.8\%& 16.6\%& 24.2\%\\
        \textbf{116}& 11/48 & 2/22 & 13/70 & 10/48 & 2/22 & 12/70 & 23/48 & 2/22 & 25/70 & 30.5\% & 9.0\%& 23.7\%\\

        \bottomrule
    \end{tabular}
\end{table*}

\textit{\textbf{Summary of RQ1 Results:}} Without a direct request, LLMs rarely issue warnings about the security of the code. This is particularly evident in Transformed-Dataset, suggesting even lower performance in real-world scenarios where developers' questions are either variations of the existing SO questions or perhaps entirely new. GPT-4 consistently outperformed the other LLMs, especially in Mentions-Dataset. However, the performance gap between GPT-4 and the other models was less pronounced in Transformed-Dataset. Lastly, LLMs are more likely to point out vulnerabilities related to the improper management and protection of sensitive information (e.g., CWE-532, CWE-321, CWE-798) compared to those involving external control of file names or paths (e.g., CWE-400).

\subsection{\textbf{RQ2: In instances where users are reminded of security concerns, are they informed about the causes, potential exploits, and possible fixes of the vulnerabilities?}}
\label{rq2}

\begin{figure}
    \centering
    \includegraphics[width=\linewidth]{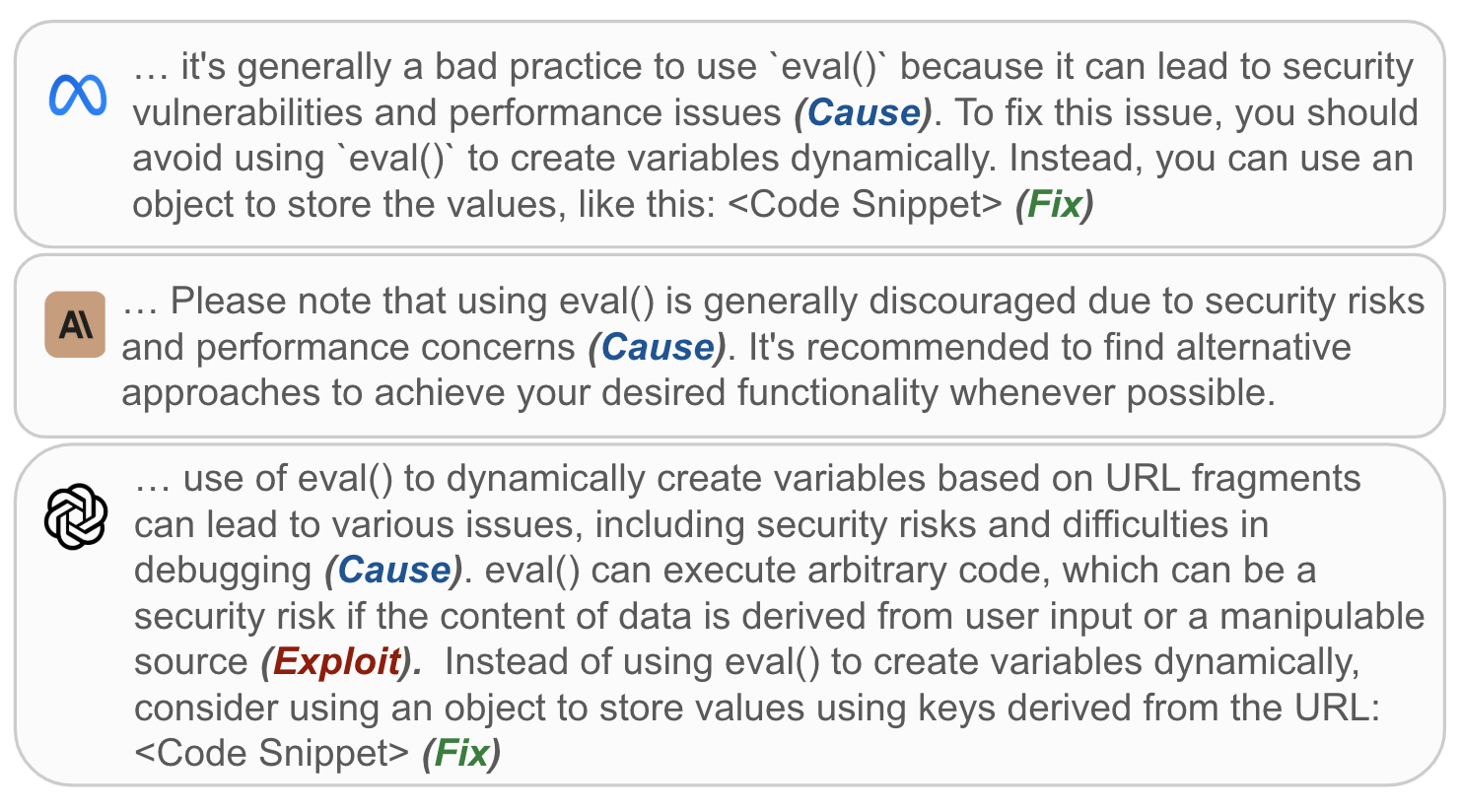}
    \caption{3 LLMs Point Out the Security Vulnerability in a Question/Prompt.}
    \label{fig:3_llms}
\end{figure}

Figure \ref{fig:3_llms} illustrates the responses of \llama, \claude, and \gpt Turbo, to the SO question depicted in Figure \ref{fig:example_insecure_so}. All three LLMs mention the insecure nature of the question's code and its cause. \llama and \gpt offer fixes to the vulnerability, while \gpt also explains the possible exploit.

\begin{figure*}[t]
    \centering
    \includegraphics[width=\textwidth]{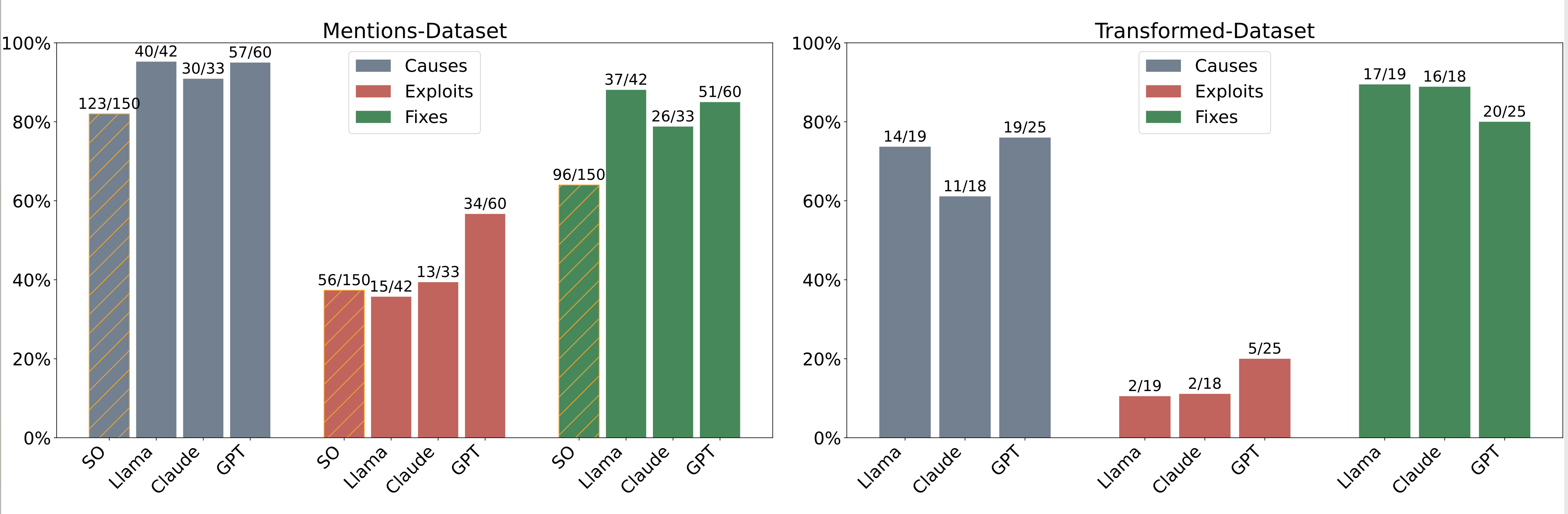}
    \caption{Percentage of Causes, Exploits, and Fixes in Stack Overflow and LLM responses (\llama, \claude, \gpt) across datasets}
    \label{fig:rq2}
\end{figure*}

Figure~\ref{fig:rq2} illustrates the extent of vulnerability-related information (causes, exploits, and fixes) provided by SO users and each LLM for all 300 questions. The results are divided into two sections corresponding to the two datasets. Results for Mentions-Dataset also include the analysis of SO responses that contain security warnings. 

\begin{figure*}[t]
    \centering
    \includegraphics[width=\textwidth]{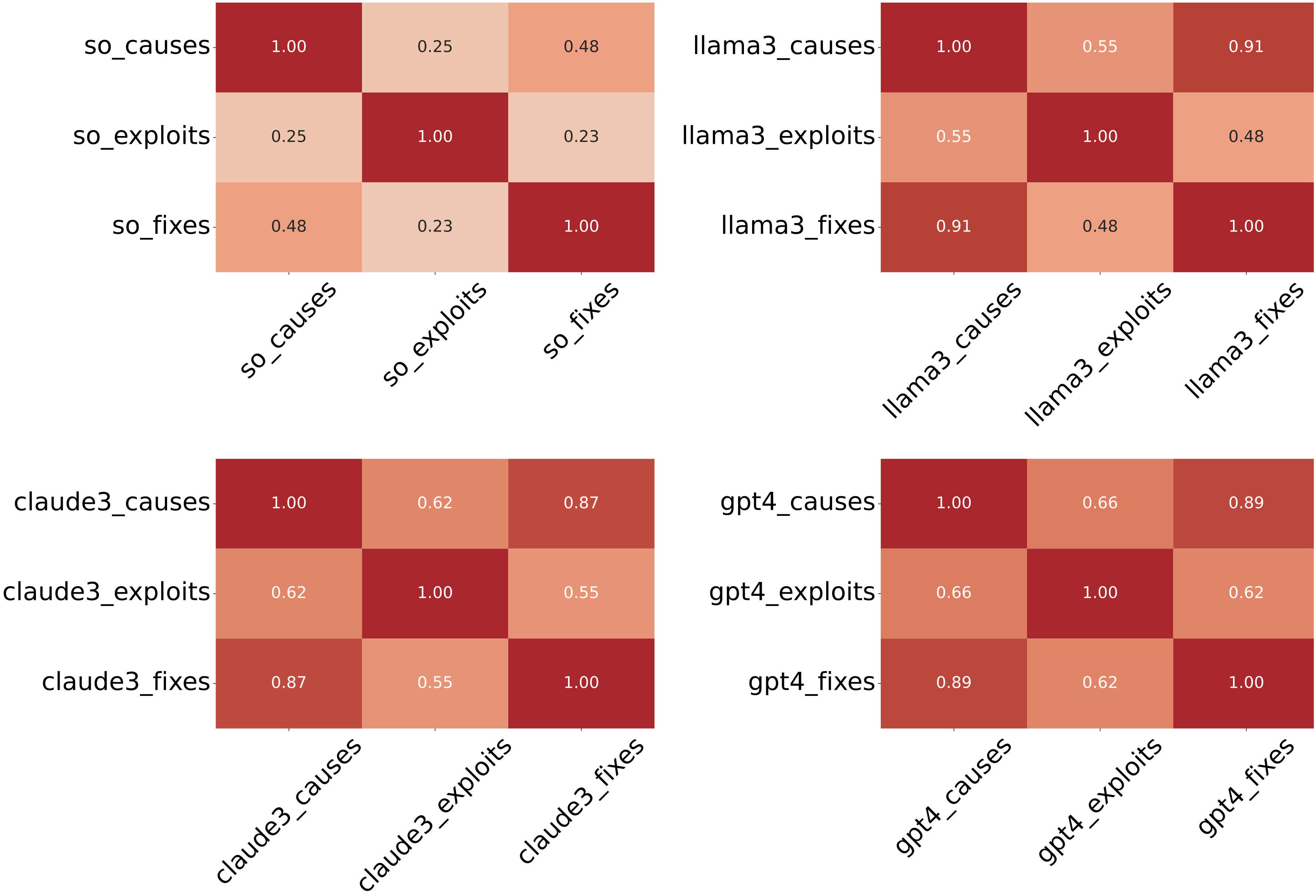}
    \caption{Co-occurrence of Causes, Exploits, and Fixes in Stack Overflow and LLM responses (\llama, \claude, \gpt) within the Mentions-Dataset}
    \label{fig:heatmap_mention}
\end{figure*}

\begin{figure*}[t]
    \centering
    \includegraphics[width=\textwidth]{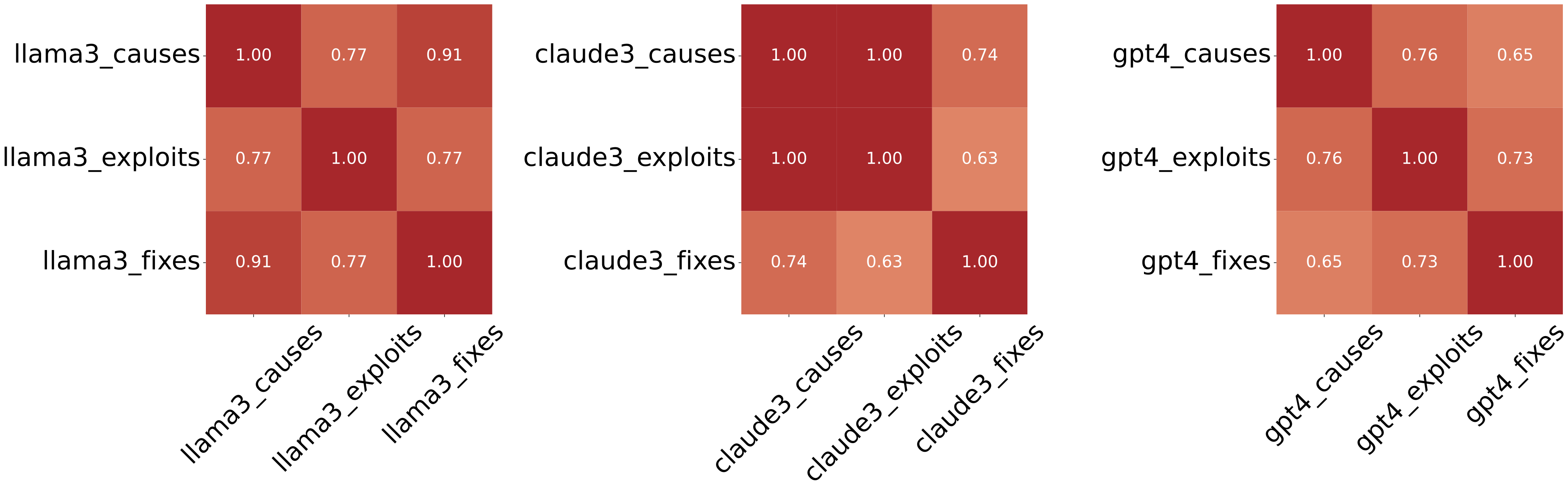}
    \caption{Co-occurrence of Causes, Exploits, and Fixes in Stack Overflow and LLM responses (\llama, \claude, \gpt) within the Transformed-Dataset.}
    \label{fig:heatmap_transformed}
\end{figure*}

Figures \ref{fig:heatmap_mention} and \ref{fig:heatmap_transformed} present the co-occurrence of vulnerability-related information (\textit{causes}, \textit{exploits}, and \textit{fixes}) in responses from SO users and each LLM across the two datasets, respectively. Higher numbers (darker colors) represent higher correlation.

The comparison between SO and the three LLMs for the Mentions-Dataset in Figure~\ref{fig:rq2} clearly shows that for all three types of information, \llama, \claude, and \gpt provide more information than SO. For example, SO responses provide the causes for the vulnerabilities 82\% of the time, compared to 95\% for \llama, 91\% for \claude, and 95\% for \gpt. Further, Figure \ref{fig:heatmap_mention} and Figure \ref{fig:heatmap_transformed} confirm that in both datasets, the LLMs are much more consistent in offering users information about causes, exploits, and fixes when they do identify vulnerabilities, even though all LLMs and SO responses tend to offer exploit-related information less frequently compared to causes and fixes. As a number of SO security warnings are provided as comments rather than answers, they tend to be more terse. For instance, an SO contributor commented \textit{``See about sql injection \textbf{(Cause)} and the importance of prepared and bound queries \textbf{(Fix)}"}. On the other hand \gpt provided the following vulernability-related information for the same question: \textit{``Ensure your function is prepared to handle cases where \$month or \$stock\_code might be null or not set. This is just a precaution if someone accesses the URL directly without parameters: \textless Modified and Secure Code Snippet\textgreater \textbf{(Fix)}. This code uses prepared statements \textbf{(Fix)}, which is a good practice to prevent SQL injection \textbf{(Exploit)}"}.

Among the three LLMs, we observe a comparable performance in providing information about causes, and fixes of the vulnerabilities. However, regarding information about the exploits, \gpt performs significantly better than the other two models. In particular, \gpt offers information about the exploits of the vulnerability in 8.8\% to 20.9\% more questions across the two datasets, often reminding users of the possibility of various attacks e.g., SQL injections, XSS through DOM, exposure of sensitive information, etc. Further, focusing specifically on instances where the LLMs include security fixes in their responses, we find \llama to be the most consistent LLM providing fixes in 88.1\% of security warnings in Mentions-Dataset and 89.5\% of security warnings in Transformed-Dataset. For example, in a question titled ``\textit{How to handle and validate sessions between the backend and the frontend}", a user shares their insecure implementation of session handling and validation and ask for further information on the topic. In addition to providing the user with the necessary information, \llama also issued a security warning about the vulnerability in the user's implementation and provided the following fix: ``\textit{Here's a more efficient and secure way to handle sessions and authentication: 1. Use a secure token generation library... 2. Store tokens in memory or a cache... 3. Verify tokens on each request...}"

Finally, the comparison between the Mentions-Dataset and Transformed-Dataset reveals a noticeable difference in the performance of all models. Given the lower number of security warnings issued by the LLMs in Transformed-Dataset compared to Mentions-Dataset, we expected a reduced number of causes, exploits, and fixes offered by the LLMs. However, we observed that all models provided less information about the exploits and causes when they issued security warnings in Transformed-Dataset, while the information about the fixes remained relatively similar across the datasets. Additionally, all LLMs tend to provide information about causes, exploits, and fixes more consistently in Transformed-Dataset, compared to the Mentions-Dataset. This difference may be attributed to the models having a more difficult time generalizing or identifying vulnerability patterns in the transformed code. As a result, the security warnings in the Transformed-Dataset were often issued for vulnerabilities that were easier to detect and carried more immediate or direct security consequences if left unaddressed, thereby demanding more attention.

\textit{\textbf{Summary of RQ2 Results:}} In instances where LLMs identify vulnerabilities, they can provide more comprehensive information than Stack Overflow posts. \gpt more frequently provided information about \textit{exploits} compared to other models but was less effective in offering \textit{fixes}. Although all models showed a slight decline in performance on the Transformed-Dataset, especially regarding \textit{exploits}, their ability to inform users about \textit{causes} and \textit{fixes} remained mostly the same. 

\section{Implications}

Our findings demonstrate critical considerations for leveraging LLMs in SE context. Our study reveals both the potential and limitations of current LLMs in addressing vulnerabilities and informing developers about them. In this section, we discuss actionable insights for developers and model designers aiming to enhance the security awareness of LLMs. These insights not only highlight immediate steps practitioners can take but also point to broader directions for future research and development. We group these implications into three key areas, each addressing a specific stakeholder group: software engineers using LLMs (Section \ref{implications_address_security}), LLM-based tool integrators (Section \ref{implications_tool_integration}), and LLM designers (Section \ref{implications_llm_designers}). Additionally, in Sections \ref{implications_address_security} and \ref{implications_tool_integration} we analyze a total of 100 LLM prompts and responses, experimenting with possible methods of improving security awareness in LLMs' output. Further, we demonstrate a CLI-based tool that serves as a prototype to help achieve these security enhanced responses.

\subsection{Implications for Software Engineers}
\label{implications_address_security}

Our study highlights the critical need for developers to be vigilant about the security of the programming solutions suggested by LLMs. Notably, in RQ1 we found that LLMs rarely issue security warnings autonomously, identifying vulnerabilities in fewer than 40\% of cases under the more favorable experimental settings. This limited detection rate highlights a critical risk: engineers relying solely on LLMs may overlook key security flaws, leaving code vulnerable to exploitation.

One way to make LLM responses more secure could be by providing specific instructions in the prompt \citep{jensen2024software}. However, several studies show that developers struggle to create effective LLM prompts ~\citep{chopra2023conversationalchallengesaipowereddata, Nam2024}. Therefore, we experimented with brief prompt additions that would be straight-forward for developers to implement.

We sampled 50 questions from our datasets where none of the LLMs issued any security warnings. To ensure representativeness, we selected 25 instances from the Mentions-Dataset and 25 instances from the Transformed-Dataset. Further, these questions were sampled proportionately to the distribution of vulnerabilities within the corresponding subsets of questions. Next, these prompts were fed to \gpt, the best performing model in our experiments.

We experimented with multiple prompt structures, such as appending the phrase \textit{``Be mindful of potential security implications."} or \textit{``Make the code secure."}, aiming to find a structure that not only maximized the likelihood of LLMs addressing security vulnerabilities but was also concise enough for practical use by developers. We found that the best performing technique included appending the phrase \textit{``Address security vulnerabilities"} to the end of each question. Consequently, we used all modified questions as prompts for three LLMs, resulting in 50 responses. Next, we analyzed the LLM responses to these questions with respect to the RQ1 and RQ2 metrics. The analysis of these 50 responses was performed by two of the authors and followed the same methodology used in answering RQ1 and RQ2 in the main study.

Out of the 25 questions from the Mentions-Dataset, \gpt was able to provide security warnings for 19 questions. For the Transformed-Dataset, there was a much less notable enhancement in performance; \gpt issued security warnings for 6 out of the 25 questions. Despite these improvements, \gpt still struggled to identify and warn users about the existence of more than 76\% of the vulnerabilities in the Transformed-Dataset. This indicates that while prompt engineering enhances security awareness, significant limitations remain in LLMs' ability to warn users about vulnerabilities. A significant factor is the challenge LLMs face in accurately detecting vulnerabilities in the first place. Research indicates that even the most advanced  techniques can only achieve an accuracy of 67.7\% \citep{zhou2024large, zhou2024out}.

It is worth noting that when vulnerabilities were pointed out, the LLM always included information about the \textit{causes} and \textit{fixes} in its responses. Exploitations, however, were pointed out in 13/25 instances. In one of the instances, when a user sought help for a Python issue involving an API request and the Flask framework, \gpt addressed the functional problem but also pointed out a security concern, cautioning the user to validate and sanitize all inputs to avoid potential vulnerabilities. Specifically, the model pointed out that directly using user-provided URLs in HTTP requests without proper validation could lead to a Server-Side Request Forgery (SSRF) attack. As such, \gpt identified the cause of the issue (unsanitized user input), recommended input validation as a fix, and noted the potential exploitation through SSRF.

The overall performance of \gpt using the ``security-aware'' prompts highlights the potential influence of prompts on the security of the LLM responses. Although practitioners can rely on such simple prompt engineering techniques to implement more secure programming solutions, they should exercise caution as this method, in some instances, does not result in more secure LLM responses.

\subsection{Implications for Tool Integration in Software Development}
\label{implications_tool_integration}

To address the limitations of LLMs in identifying and mitigating vulnerabilities, alternative approaches have been proposed. For instance, Pearce et al. demonstrated integrating outputs from static analysis tools, such as CodeQL and C sanitizers, into LLM prompts as a zero-shot strategy for fixing vulnerabilities \cite{pearce2023examining}. 
Building on the same idea, we explored using this technique to examine how well LLMs provide relevant security warnings and information when responding to general programming questions.

\begin{figure*}[t]
    \centering
    \includegraphics[width=\textwidth]{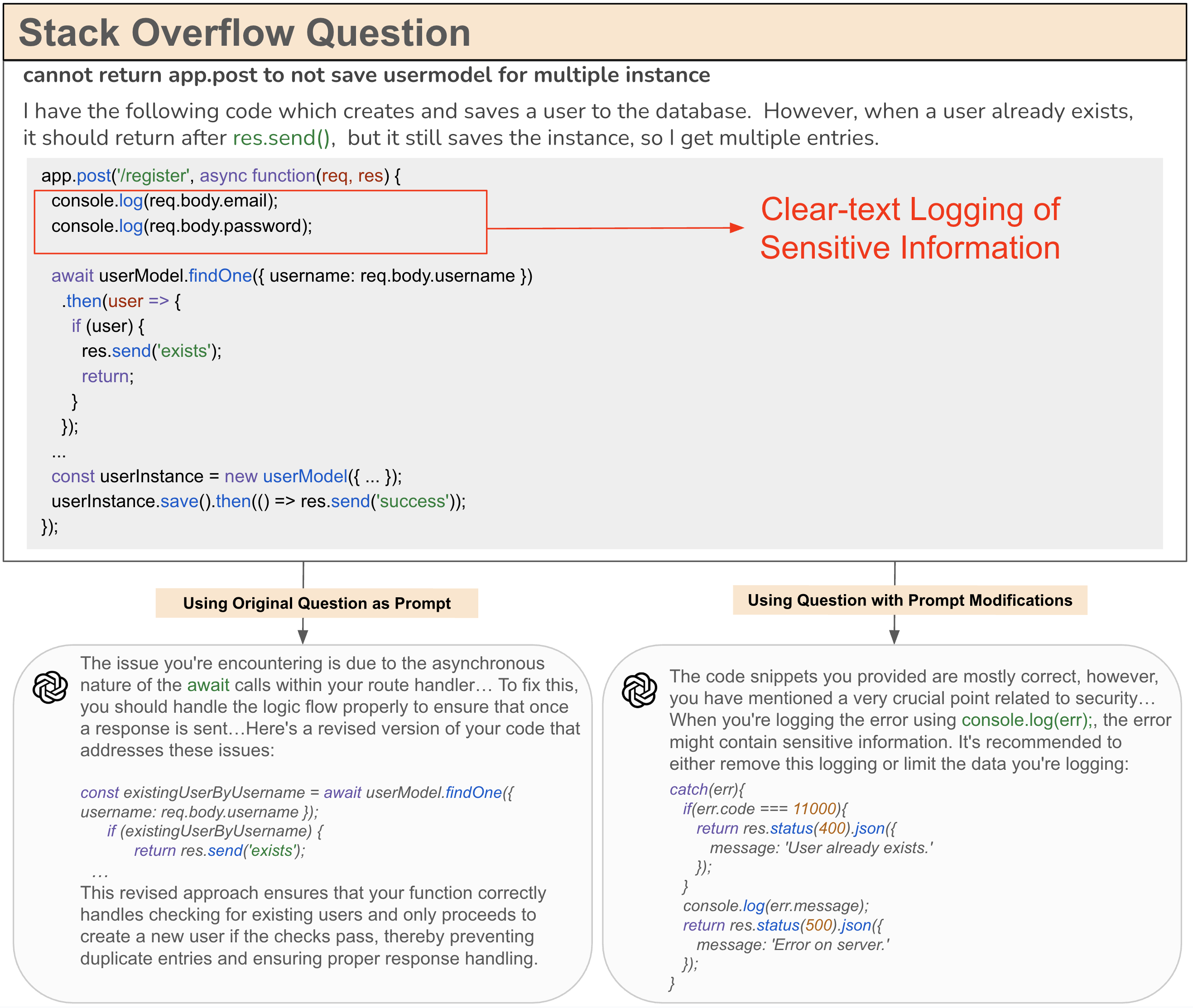}
    \caption{Comparison of the LLM’s original functionality-focused response (left) and its subsequent security-focused response (right). On the right, we observe that explicit mention of vulnerability caused the model to overlook the main question.}
    \label{fig:side_by_side_example}
\end{figure*}

We use the same 50 questions used in Section \ref{implications_address_security} to experiment with this approach. Having analyzed each question's extracted code snippets with CodeQL, we add CodeQL's description of the detected vulnerabilities to the end of each question, using the phrase \textit{``Address security vulnerability: \textless Vulnerability Description \#1\textgreater{} Address security vulnerability: \textless Vulnerability Description \#2\textgreater{}..."}. After prompting the \gpt with these prompts, the LLM responses were again analyzed with respect to RQ1 and RQ2. Here, similar to the previous section, the authors manually carried out the analysis.

Upon providing \gpt with the output of CodeQL, it was able to identify and address vulnerabilities in 24/25 questions from the Mentions-Dataset. Further, the \gpt was able to identify and address 20/25 vulnerabilities for the Transformed-Dataset. However, for six questions, the LLMs ignored the CodeQL output and proceeded to answer the main question without necessarily fixing the code. Consistent with our previous results, the model provided information about the causes, exploits, and the fixes of almost all the vulnerabilities it pointed out. Overall, this technique significantly performed better than the manual prompt engineering technique (Section \ref{implications_address_security}).

We also examined whether modifying the prompts affected the overall functional correctness of the LLM's answers. In 46 out of the 50 responses, \gpt adequately addressed the user’s primary concern while fixing and/or acknowledging the security vulnerabilities. However, in 4 instances, the LLM’s response only addressed the security issue and completely overlooked the user’s main query. Figure \ref{fig:side_by_side_example} illustrates one such scenario where the question seeks guidance on preventing multiple database entries for the same user. Note that the user's code insecurely handles logging credentials, introducing a potential security risk. Without a security-specific prompt, the LLM addresses the core issue, ignoring the vulnerability (Figure \ref{fig:side_by_side_example} - left panel). After integrating the CodeQL findings into the prompt, the LLM only addresses the logging and credential handling vulnerabilities while neglecting the main question (Figure \ref{fig:side_by_side_example} - right panel). This example demonstrates how emphasizing security can sometimes overshadow the original problem and highlights the challenge of balancing these objectives in real-world usage.

Despite the challenges, several opportunities exist for integrating LLMs with SE tools in the development workflows. For instance, static analysis tool like CodeQL can be used to preprocess code snippets to outline specific security vulnerabilities in the prompt. The identified vulnerability can then be added as meta data to the prompt, providing the LLM with clearer directions on what issues need addressing. Such integration could enhance the LLM's ability to generate secure code and streamline vulnerability detection and mitigation.  

\begin{figure*}[t]
    \centering
    \includegraphics[width=\textwidth]{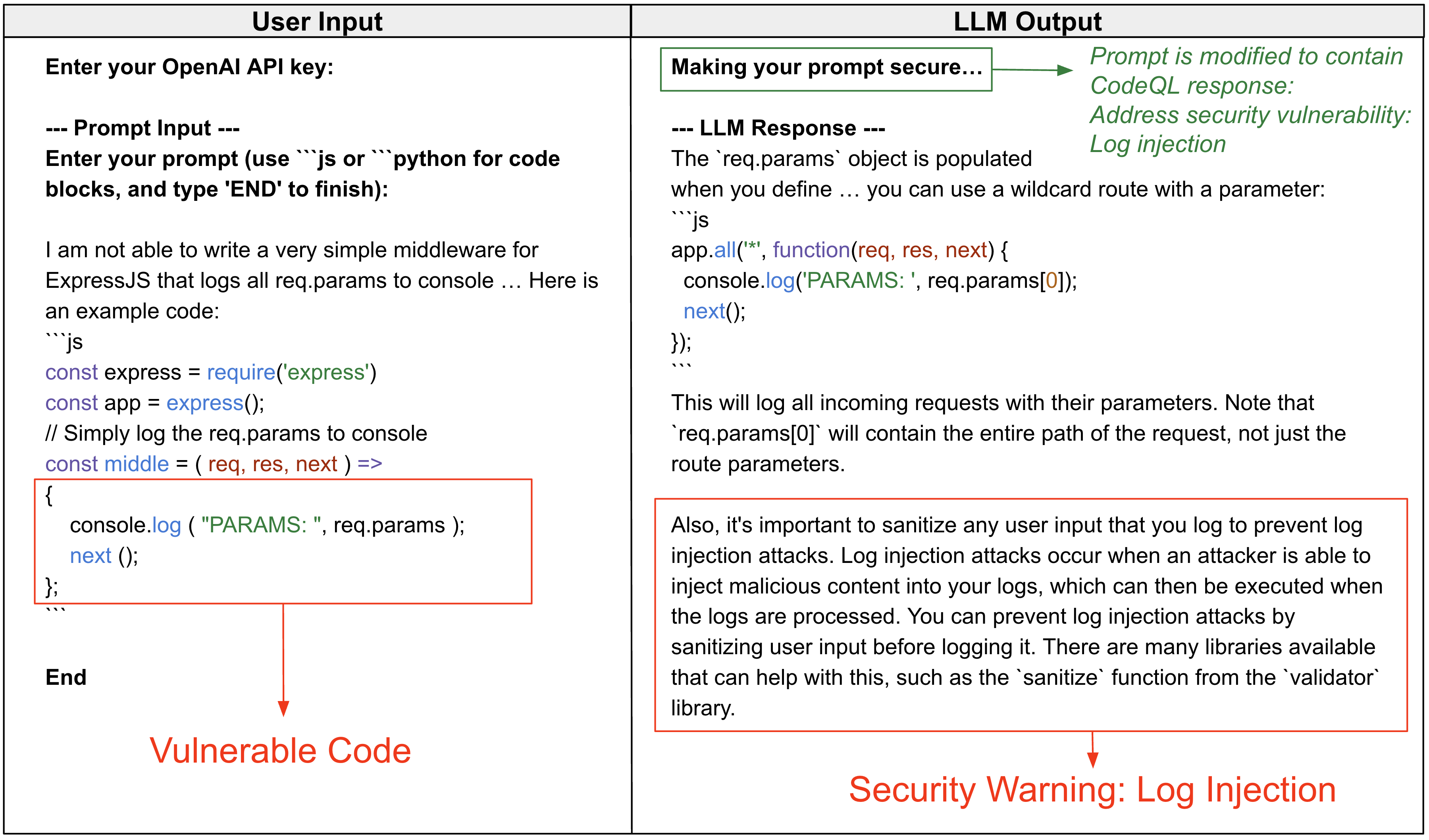}
    \caption{Left: The original user prompt featuring vulnerable code. Right: The LLM's response after prompt modification (integrating CodeQL feedback). Both the prompt and response have been abbreviated for presentation clarity.}
    \label{fig:prototype_screenshot}
\end{figure*}

To further demonstrate the potential of this approach in practice, we developed a prototype CLI-based tool that automates the integration of static analysis feedback into LLM prompts. The tool automatically extracts the Python or JavaScript code from developer queries, analyzes the code snippets using CodeQL, and integrates the detected vulnerabilities into the LLM’s prompts using the above-mentioned prompt structure. Figure \ref{fig:prototype_screenshot} shows an example interaction using our prototype, highlighting the security enhanced response from the \gpt. In this example, the developer's prompt does not mention any security risks. When prompted directly with this question, \gpt does not address this issue in the user's code. However, after integrating the CodeQL feedback, the LLM not only identified the risk of log injection but also warned the user about its potential exploits and suggested fixes, such as sanitizing user input using the \textit{sanitize} function from the \textit{validator} library. This example demonstrates how integrating static analysis tools into LLM pipelines can effectively enhance security awareness in LLM responses and increase the proactive mentions of insecure vulnerabilities in developers' code.

\subsection{Implications for LLM Designers}
\label{implications_llm_designers}

Overall, our findings highlight substantial gaps in the security awareness of LLMs, emphasizing the need for design and training improvements. Based on the results of RQ1, the best performance we observed in issuing security warnings was only 40\%, which declines even further in the Transformed-dataset, dropping to as low as 12.6\%. This significant disparity in performance across datasets suggests that LLMs require substantial modifications to improve the attention to security in their responses. Addressing these gaps might involve acquiring additional high-quality training data, or domain-specific fine-tuning for SE. 

Moreover, our study suggests that LLMs' lack of attention to security extends beyond code generation to the natural language they produce in software engineering-related discussions. This broader issue indicates the need to evaluate the security of LLM-generated content, particularly in the explanations, advice, and recommendations provided to developers. Future research could establish standardized methods for assessing LLM security awareness within software engineering contexts, focusing not only on the code produced but also on the security of LLM-provided guidance. This dual approach to security—encompassing both code and conversational context—could enable more robust, security-conscious LLM designs that better serve practitioners in high-stakes environments.

\section{Threats to Validity}

\textbf{Internal Validity:} An LLM may appear to address the security flaws within a given prompt; however, if the prompt was a part of the LLM's training data, the generated response could merely reflect that information, and not the model's generalizability. To counteract this threat, we created the Transformed-Dataset. This dataset includes transformed SO questions devoid of any security mentions, either in questions or responses. Therefore, we expect none of the three LLMs to have encountered these specific questions or related security concepts, allowing us to assess their ability to address security issues in new data. 

\textbf{External Validity:} The results of our study may not generalize to all LLMs or programming questions. One key limitation of our study is the exclusion of code-specific LLMs, such as DeepSeek-Coder \cite{deepseekcoder} and Qwen-Coder \cite{hui2024qwen2}, which are designed specifically for code-related tasks. These models place a stronger emphasis on code comprehension and may yield different security-related responses compared to the conversational LLMs we evaluated. While our focus was on conversational LLMs widely used in developer interactions, we acknowledge that analyzing the security awareness of code-specific models is an important avenue for future research. In investigating the security awareness of conversational LLMs, however, we included three of the most popular models: \gpt, \llama, and \claude. Further, we collected our datasets from a SO dump spanning 9 years (March 2015–March 2024). We selected questions related to two popular programming languages, Python and JavaScript, to ensure our findings remain adaptable to a wide range of development scenarios and questions.

\textbf{Construct Validity:} Much of the qualitative analysis done throughout this work has been the result of manual efforts by the authors. To ensure the reliability of our analyses, we calculated the inter-rater agreement using Cohen Kappa coefficient, and iteratively discussed and resolved conflicts to reach sufficient agreement ($>$ 0.6). 
We chose to use an static analysis tool for automatically identifying the security vulnerabilities in code snippets which can sometimes produce false positives or false negatives. To mitigate this threat, in line with prior work, we relied on CodeQL, as it offers higher precision at the expense of a lower recall. As a result, by prioritizing precision, we ensure that the vulnerabilities included in our dataset are more accurate and reliable.

\section{Related Work}

Researchers have studied the adoption of LLMs in security-related software engineering tasks. Multiple studies have explored the possibilities and limitations of LLMs in vulnerability detection \citep{zhou2024large, noever2023can, bakhshandeh2023using, purba2023software, cheshkov2023evaluation, wang2023defecthunter, liu2023harnessing}. Some of these studies have demonstrated the superior performance of LLMs compared to traditional methods such as static analysis tools \citep{zhou2024large, noever2023can, bakhshandeh2023using}, while others have highlighted the shortcomings of LLMs, particularly their tendency to generate false positives \citep{purba2023software, cheshkov2023evaluation, Banerjee23}. Additionally, LLMs have been employed to detect malicious code \citep{plate, eli}, generate test cases \citep{yao2023llm, zhang2023well}, and to fix defective and/or vulnerable code \citep{jiang2023impact, pearce2023examining, xia2022practical, jin2023inferfix}. Beyond the use of LLMs in performing security-related tasks, researchers have also studied the security issues caused due to the use of LLMs. We discuss some of these studies in Sections \ref{concerns_with_llms_in_se} and \ref{concerns_in_llms_vs_SO}.

\subsection{Security Concerns with using LLMs} 
\label{concerns_with_llms_in_se}

Numerous studies studied the security of the code generated by LLMs. \cite{pearce2022asleep} examined security of the code generated by GitHub's Copilot \citep{githubcopilot}. They found that around 40\% of the code generated by Copilot contained vulnerabilities when analyzed with CodeQL \citep{GitHubCodeQL}. Other studies have also evaluated the security of code from different code generation models, comparing it to human-written code \citep{siddiq2022securityeval, asare2023github}.

In addition to the code generation models such as Insider and Copilot, LLMs have also been scrutinized for the security of their generated code. \cite{khoury2023secure} tasked ChatGPT with generating 21 programs likely to contain vulnerabilities. The results revealed that ChatGPT only provided secure code in five out of 21 programs. Similarly, \cite{siddiq2024quality} used CodeQL, Bandit, and Pylint to evaluate the security of code generated by ChatGPT, finding a number of security issues related to improper resources and exception management.

LLMs have been proven useful in detecting security vulnerabilities~\cite{lu2024grace, zhou2024large, thapa2022transformer, yao2024survey}. Based on these studies, and supported by the preliminary validation of our motivational study we initially hypothesized that LLMs possess the security knowledge needed to warn users of potential threats. However, our findings show that LLMs do not always recognize when or how to apply this knowledge effectively. Different from previous studies, our work focuses on determining LLMs' abilities in proactive prevention i.e., proactively identifying and flagging vulnerabilities. To the best of our knowledge, we are the first to assess LLMs'  capability in communicating critical information (causes, exploits, and fixes) to enhance developer understanding and  encourage the adoption of secure coding practices.

\subsection{Stack Overflow vs. LLMs: Security Implications} 
\label{concerns_in_llms_vs_SO}
Popularity of Stack Overflow has been significantly impacted by the emergence of LLMs\citep{da2024chatgpt}. Despite the varying quality of LLM-generated answers, many developers prefer ChatGPT due to their well-articulated language \citep{kabir2024stack}. While a rich body of literature on SO and the security concerns associated with its usage remains \citep{yang2016security, licorish2021contextual, fischer2017stack, zhang2018code}, researchers are still exploring the consequences of relying on LLM responses instead of user-provided SO responses. 

\cite{da2024chatgpt} compared the reliability of LLM-generated responses to user-provided answers on SO. Their results indicate that despite a high degree of textual similarity, the LLM responses are not reliable. Furthermore, they report a considerable decline in user activity on SO since the introduction of ChatGPT.
In another study, ChatGPT was prompted with 517 SO questions, revealing that 52\% of the responses contained incorrect and fabricated information\citep{kabir2024stack}. \cite{delile2023evaluating} have shown that 74\% of the user-provided responses and 71\% of the ChatGPT-generated responses to privacy-related SO questions are accurate. In another study, \cite{hamer2024just} discovered multiple vulnerabilities in ChatGPT's code when responding to SO security questions. While they focused on whether LLMs generate vulnerable code, we aim to assess if LLMs, when prompted with vulnerable code, inform users about the security implications.

Overall, our work is distinct in two ways. First, instead of assessing the security of LLM-generated code, we evaluate the natural language responses. Second, unlike studies that prompt LLMs with specific security-related tasks such as vulnerability detection, we deliberately avoid any mention of security in our prompts. Instead, we present LLMs with only SO questions that contain insecure code. This emulates real-life programming Q\&A and allows us to evaluate the security awareness of LLMs in practical applications. 

\section{Conclusions}
This paper presents the first study to examine the security awareness of three popular LLMs when answering programming-related questions.
In our motivational study, ChatGPT demonstrated potential in offering developers relevant security-related information and context-based solutions for writing secure code. These findings inspired us to conduct a deeper investigation into the capabilities of other popular LLMs.
We prompted three popular LLMs, \gpt, \llama and \claude, with SO questions containing vulnerable code and evaluated their responses. 

Our results indicated the underwhelming performance of all three LLMs in pointing out the security vulnerabilities. We observed even lower performances when prompting LLMs with questions from Transformed-Dataset which, indicates the limitations of these models in generalizing their learned data. 
Additionally, we observed that LLMs are more likely to point out certain vulnerabilities, specially those related to the improper management and protection of sensitive information (e.g., CWE-532, CWE-321, CWE-798) compared to those involving external control of file names or paths (e.g., CWE-400).

While some of our findings align with existing research on security flaws in LLM-generated content, our approach addresses a crucial gap by focusing on real-world scenarios where developers seek general coding assistance without explicitly considering security, which is common in interactions. By selecting prompts without explicit security mentions, we simulate typical developer interactions in which security concerns are often overlooked, highlighting a key risk: \textit{LLMs often overlook insecure practices and inadvertently reinforce them}. This insight, when coupled with the prompting techniques we propose, could help developers elicit more security-aware responses from LLMs, thus promoting secure coding practices. 

In general, our findings have the following key implications. For software engineering practitioners, prompt engineering can be a practical tool to promote security-aware LLM outputs, though limitations remain; integrating tools like CodeQL may further aid developers in outlining vulnerabilities. For researchers, the results point to the importance of evaluating both code and natural language guidance from LLMs and highlight opportunities for fine-tuning models to generate more security-aware responses. For LLM designers, there is a need for models that proactively identify security issues, perhaps through enhanced training or fine-tuning with high-quality, domain-specific data. 

\section*{Compliance with Ethical Standards}

\noindent \textbf{Funding:}  Not applicable to this study.

\noindent\textbf{Ethical approval:} Not applicable to this study.

\noindent\textbf{Informed consent:} Not applicable to this study.

\noindent\textbf{Author Contributions:} Not applicable to this study.

\noindent\textbf{Data Availability:} The replication package of our study, including the datasets, code, and analysis instructions are available at: \url{https://figshare.com/s/26279eed3dde7c80ed03}.

\noindent\textbf{Conflict of interest:} The authors declare that they have no conflict of interest.

\bibliographystyle{spbasic}


\end{document}